\documentclass[12pt]{iopart}
\usepackage{graphics}
\usepackage{bm}
\usepackage{cite}

\begin{document}
\bibliographystyle{prsty}

\input{psfig} 

\title[Multiple Projection Optical Tomography]{Multiple Projection
  Optical Diffusion Tomography with Plane Wave Illumination}
 
\author{Vadim A. Markel\footnote[1]{vmarkel@mail.med.upenn.edu}
  and John C. Schotland\footnote[2]{schotland@seas.upenn.edu}}

\address{Departments of Radiology and Bioengineering, University
  of Pennsylvania, Philadelphia, PA 19104}

\date{\today} 

\begin{abstract}
  We describe a new data collection scheme for optical diffusion
  tomography in which plane wave illumination is combined with
  multiple projections in the slab imaging geometry. Multiple
  projection measurements are performed by rotating the slab around
  the sample.  The advantage of the proposed method is that the
  measured data can be much more easily fitted into the dynamic range
  of most commonly used detectors. At the same time, multiple
  projections improve image quality by mutually interchanging the
  depth and transverse directions, and the scanned (detection) and
  integrated (illumination) surfaces. Inversion methods are derived
  for image reconstructions with extremely large data sets.  Numerical
  simulations are performed for fixed and rotated slabs.
\end{abstract}

\pacs{87.57.Gg,42.30.Wb}
\submitto{\PMB}
\maketitle

\section{Introduction}
\label{sec:intro}

Tomographic imaging with diffuse light, often referred to as optical
diffusion tomography (ODT), is a novel biomedical imaging
modality~\cite{arridge_99_1,boas_01_1}. Although ODT was introduced
more than a decade ago, efforts to bring it into the clinical
environment are hampered by relatively low quality and spatial
resolution of images. Therefore, optimization of image reconstruction
algorithms for high-resolution ODT is of fundamental importance. In
this paper we study the image reconstruction problem of ODT by
combining three novel approaches.  First, we employ analytic image
reconstruction methods which allows the utilization of extremely large
data sets~\cite{markel_02_1,markel_03_1}. Second, we make use of
multiple projections~\cite{markel_04_2}. Here by multiple projections
we mean multiple orientations of the measurement apparatus with
respect to the medium. Finally, we utilize the recently proposed plane
wave illumination scheme~\cite{xu_alfano_01_1}.  Each of these methods
provides an advantage which is not lost when the techniques are
combined. We begin by briefly reviewing the approaches to ODT imaging
mentioned above. Note that throughout this paper we consider the slab
imaging geometry which is often used in mammography and small-animal
imaging~\cite{franceschini_97_1,ntziachristos_99_1}. In order to
obtain multiple projection measurements, a pair of parallel plates are
rotated around the medium to be imaged which is assumed to be
stationary and unperturbed.

There is a direct relationship between the spatial resolution of
images and the number of data points used for
reconstruction~\cite{markel_02_1}. Indeed, the reconstruction of an
image with $N$ voxels, in principle, requires at least $N$
measurements. In practice, the ill-posedness of the image
reconstruction problem and the presence of noise require that this
number be larger than $N$.  Measurements with up to $10^{10}$ data
points are feasible with CCD camera-based instruments.  However, many
previous studies of the image reconstruction problem in ODT have been
limited to relatively small data sets (e.g., 256 data points in
Ref.~\cite{pogue_99_1}, 900 data points in Ref.~\cite{culver_01_1}).
This can be explained by the high computational complexity of
algebraic image reconstruction algorithms which scales as $O(N^3)$. To
ameliorate this difficulty, we have recently introduced a family of
analytic image reconstruction algorithms that can utilize extremely
large data
sets~\cite{schotland_97_1,schotland_01_1,markel_01_3,markel_02_2,markel_04_4}.
These methods allow a dramatic reduction in computational complexity
which, in turn, leads to a significant improvement of spatial
resolution of images. However, these methods have certain limitations.

First, the data collection method described in Ref.~\cite{markel_02_2}
requires that measurements are taken for source-detector pairs
separated by a distance which is much larger than the slab thickness.
In practice, such measurements are technically difficult to perform.
Reduction of the required dynamic range of the detectors can be
achieved by using {\it plane wave illumination}~\cite{xu_alfano_01_1}.
Note that due to the general theoretical reciprocity of sources and
detectors, plane wave illumination and scanned detection is equivalent
to integrated detection and scanned narrow beam illumination. However,
in a practical situation, the different nature of illuminating and
detecting devices must be taken into account. For the sake of
definitiveness, we consider below plane wave illumination and combine
it with analytic image reconstruction methods. Note that plane
wave illumination requires time- or frequency-resolved measurements.
However, it can be seen that the number of degrees of freedom in the
data is still insufficient for unique, simultaneous reconstruction of
the absorption and diffusion (or reduced scattering) coefficients.
This situation is similar to the nonuniqueness demonstrated in
Ref.~\cite{arridge_98_1}.  Therefore, we focus here on the
reconstruction of absorbing inhomogeneities assuming that the
diffusion coefficient of the medium is constant. Reconstruction of
purely absorbing inhomogeneities have been employed, for example, in
breast imaging~\cite{colak_99_1,hawrysz_00_1,culver_03_1,intes_03_1}
or blood oxygenation level imaging~\cite{vanhouten_96_1,culver_03_3}.

Second, it was shown in Ref.~\cite{markel_02_1} that in the slab
imaging geometry the depth resolution (in the direction perpendicular
to the slab) is fundamentally different from the transverse resolution
(in the direction parallel to the slab surface). The depth resolution
is much more sensitive to noise and the point-spread functions (PSFs)
in the depth direction strongly depend on the location of the
inhomogeneity. This results in image artifacts. In general, the
non-uniformity of the PSF can be a serious problem if more than one
inhomogeneity is present. To correct this situation, we have recently
proposed multi-projection image reconstruction
methods~\cite{markel_04_2,markel_04_4}. Multiple projections render
the depth and transverse directions mutually interchangeable.  As a
result, the PSF becomes more uniform and less position-dependent,
and also more sharply peaked. Note that multiple projections have been
used in X-ray imaging for some time. However, an important difference
between ODT and X-ray computed tomography is that, in the first case,
tomographic imaging is possible {\it in principle} with a single
projection while in the second case it is not.  Perhaps, due to this
fact, multiple projections in optical tomography have not been
investigated until recently, except for the case of ballistic
propagation without scattering (e.g.~\cite{brown_92_1}), or in
conjunction with a modified version of X-ray backprojection tomography
with phenomenological corrections introduced to compensate for
scattering~\cite{colak_97_1,matson_99_1}. In Ref.~\cite{markel_04_4}
we have developed a general theoretical formalism for inverting
measurements obtained from multiple projections. In
Ref.~\cite{markel_04_2} image reconstruction with two orthogonal
projections was numerically implemented.

In this paper we implement the more general image reconstruction
algorithm of Ref.~\cite{markel_04_4} for treatment of more than two
projections in conjunction with plane wave illumination. Note that the
plane wave illumination is advantageous when measurement are
limited by the dynamic range of detectors. If the dynamic range is not
an important experimental factor, the traditional measurement scheme
with point sources and point detectors is expected to provide superior
image quality.  We combine the advantageous features of these two
approaches with the computational efficiency of the analytic image
reconstruction methods.

\section{Theory}
\label{sec:theory}

\subsection{Single projection}
\label{subsec:single_proj}

We assume that propagation of multiply-scattered light in tissue is
described by the diffusion equation. In addition, we will work in the
frequency domain with the sources harmonically modulated at the
frequency $\omega$ and detectors which yield the oscillatory part of
transmitted intensity. Then the density of electromagnetic energy in
the medium $u({\bm r})$ obeys the diffusion equation

\begin{equation}
\label{diff_eq}
- D_0 {\bm \nabla}^2 u({\bf r}) + [\alpha({\bf r}) - i\omega] u({\bf
  r}) = S({\bf r}) \ , 
\end{equation}

\noindent
where $\alpha({\bm r})$ is the position dependent absorption
coefficients, $S({\bm r})$ is the source function and the $D_0$ is the
diffusion coefficient.

Consider a slab of thickness $L$ with the plane of incidence located
at $x=-L/2$ and the detection plane at $x=L/2$. The medium is located
in the region $-L/2<x<L/2$. If point-like sources and detectors are
used (typically, thin optical fibers), the data can be expressed as a
function $\phi(\omega, {\bm \rho}_s, {\bm \rho}_d)$, where ${\bm
  \rho}_s$ and ${\bm \rho}_d$ are two-dimensional vectors specifying
the location of the sources and detectors, respectively, on the slab
surfaces. Using the first Born approximation, we linearize the forward
model by decomposing the absorption function $\alpha({\bm r})$ into a
constant background and a small fluctuating part, $\alpha({\bm r}) =
\alpha_0 + \delta \alpha({\bm r})$. We seek to reconstruct the values
of $\delta\alpha({\bm r})$ from the data $\phi(\omega, {\bm \rho}_s,
{\bm \rho}_d)$.  The usual mathematical formulation of the ODT inverse
problem is based on the integral equation~\cite{gonatas_95_1}

\begin{equation}
\label{Eq1}
\phi(\omega, {\bm \rho}_s, {\bm \rho}_d) = \int \Gamma (\omega, {\bm
  \rho}_s, {\bm \rho}_d; {\bm r}) \delta \alpha({\bm r}) d^3 r \ ,
\end{equation}

\noindent
where

\begin{eqnarray}
\Gamma (\omega, {\bm \rho}_s, {\bm \rho}_d; {\bm r}) = && \int
{{d^2q_sd^2q_d} \over {(2\pi)^4}} 
\kappa(\omega, {\bm q}_s, {\bm q}_d; x) \nonumber \\
&& \times \exp\left[ i{\bm q}_s\cdot\left({\bm \rho}-{\bm \rho}_s\right) + 
i{\bm q}_d\cdot\left({\bm \rho}_d-{\bm \rho}\right) \right] \ ,
\label{Eq2}
\end{eqnarray}

\noindent
${\bm \rho}$ is the transverse part of the vector ${\bm r}$ (${\bm
  r}=(x,{\bm \rho})$) and the form of $\kappa(\omega, {\bm q}_s, {\bm
  q}_d; x)$ is determined from the boundary conditions on the surfaces
of the slab and the expression which relates the measurable intensity
to the energy density $u({\bm r})$. The derivation of
(\ref{Eq1}),(\ref{Eq2}) and explicit expressions for $\kappa$ are
given in Ref.~\cite{markel_02_1}.  Note that the general form of
(\ref{Eq1}),(\ref{Eq2}) follows from the symmetry of the problem and
is independent of the diffusion approximation.

Next, we introduce the plane wave illumination scheme.  Instead of
using point sources located at points ${\bm \rho}_s$, we illuminate
the slab with a normally incident wide homogeneous beam of
sufficiently large diameter (compared to transverse dimensions of the
slab). At the same time we utilize point detectors. This ensures that
the new data function $\psi(\omega, {\bm \rho}_d)$ defined by

\begin{equation}
\label{psi_phi_def}
\psi(\omega, {\bm \rho}_d) = \int \phi(\omega, {\bm \rho}_s, {\bm
  \rho}_d) d^2 \rho_s 
\end{equation}
 
\noindent
has the same number of degrees of freedom as the unknown
$\delta\alpha({\bm r})$ (two spatial directions and the frequency
$\omega$).  Thus, the inverse problem is well determined. The integral
equation (\ref{Eq1}) can now be transformed to

\begin{equation}
\label{Eq3}
\psi(\omega,{\bm \rho}_d) = \int {{d^2q} \over {(2\pi)^2}}
\kappa(\omega, 0, {\bm q}; x) \exp[i{\bm q}\cdot({\bm \rho}_d - {\bm
  \rho})] \delta\alpha({\bm r})d^3r \ .
\end{equation}

\noindent
If $\psi$ is measured for $N$ different modulation frequencies and the
sources are placed on a square lattice with step size $h$,
Eq.~(\ref{Eq3}) can be inverted using the methods described
in~\cite{markel_02_2}. The SVD pseudo-inverse solution is given by

\begin{equation}
\hspace{-1cm} \delta\alpha({\bm r}) = h^2 \int_{\rm FBZ} {{d^2u} \over {(2\pi)^2}}
\exp(-i{\bm u}\cdot{\bm \rho}) \sum_{\omega,\omega^{\prime}}
P^*(\omega,{\bm u}; {\bm r}) 
\langle \omega \vert M^{-1}({\bm u}) \vert \omega^{\prime}\rangle 
\tilde{\psi}(\omega^{\prime},{\bm u}) \ .
\label{Eq4}
\end{equation}

\noindent
Here the vector ${\bm u}$ is in the first Brillouin zone (FBZ) of the
lattice of sources, namely, $-\pi/h < u_{y,z} \leq \pi/h$ and

\begin{equation}
\label{P_def}
P(\omega, {\bm u}; {\bm r})=\sum_{\bm v}
\kappa(\omega,0,{\bm u}+{\bm v}; x)\exp(i{\bm v}\cdot{\bm \rho}) \ ,
\end{equation}

\noindent
where ${\bm v}$ are reciprocal lattice vectors of the form ${\bm
  v}=(2\pi/h)(n_y\hat{\bm y} + n_z\hat{\bm z})$. The elements of
matrix the $M({\bm u})$ are given by

\begin{equation}
\label{Eq5a}
\langle \omega \vert M({\bm u}) \vert \omega^{\prime} \rangle =
\sum_{\bm v} M_1({\bm u} + {\bm v}) \ ,
\end{equation}

\noindent
where 

\begin{equation}
\langle \omega \vert M_1({\bm q}) \vert \omega^{\prime} \rangle = 
\int_{-L/2}^{L/2} \kappa(\omega,0,{\bm q};x)
\kappa^*(\omega^{\prime},0,{\bm q};x) dx
\label{Eq5b}
\end{equation}

\noindent
(the inverse matrix $M^{-1}({\bm u})$ must be appropriately
regularized~\cite{natterer_book_86}) and the Fourier transformed data
function $\tilde{\psi}(\omega,{\bm u})$ is defined as

\begin{equation}
\label{a1}
\tilde{\psi}(\omega,{\bm u}) = \sum_{{\bm \rho}_d}
\psi(\omega,{\bm \rho}_d) \exp(i{\bm u}\cdot{\bm \rho}_d) \ .
\end{equation}

\noindent
Note that, if $\delta\alpha$ is reconstructed only at points which are
commensurate with the lattice of sources, the factor $\exp(i{\bm
  v}\cdot{\bm \rho})$ is equal to unity and the function $P$ becomes
independent of ${\bm \rho}$. Note also that $\kappa$ and $M_1$ can be
calculated in terms of elementary functions~\cite{markel_04_4}.

\subsection{Multiple projections}
\label{subsec:multi_proj}

\begin{figure}
\centerline{\psfig{file=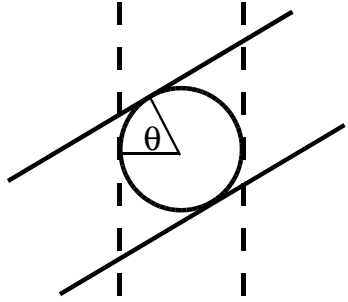,height=6cm,bbllx=0bp,bblly=0bp,bburx=252bp,bbury=288bp,clip=t}}
\caption{\label{fig:sketch}
  A sketch of the experimental set up with rotating slab.  The axis of
  rotations ($Oz$) is perpendicular to the plane of the figure and
  coincides with the axis of the cylinder $R<L/2$ inside which
  reconstructions are performed. Locations of sources and detectors
  are given in a local reference frame which rotates together with the
  slab.}
\end{figure}

We now consider inclusion of multiple projections. Let the sources and
detectors be rotated around the sample as illustrated in
Fig.~\ref{fig:sketch}. We assume that the rotations do not disturb the
medium inside the cylinder $\sqrt{x^2+y^2}<L/2$ and that the unknown
function $\delta\alpha$ vanishes outside the same region.  The space
inside the slab but outside the above cylindrical region is assumed to
have the background values of the coefficients $\alpha_0$ and $D_0$.
Experimentally, this can be implemented, for example, by rotating an
imaging apparatus around a sample suspended in matching fluid.  We
introduce cylindrical coordinates ${\bm r}=(R,z,\varphi)$ with the
$z$-axis being the axis of rotation. If the data are measured for
$N_{\theta}$ different orientations, where the respective angles
$\theta_n$ are equally spaced and given by
$\theta_n=2\pi(n-1)/N_{\theta},\ n=1,\ldots,N_{\theta}$, the
reconstruction formula (\ref{Eq4}) can be generalized
to~\cite{markel_04_4}:

\begin{eqnarray}
\hspace{-2cm}\delta\alpha({\bm r}) = {{2\pi h^2} \over {N_{\theta}}} \sum_{n=1}^{N_{\theta}}
\int_{-\pi/h}^{\pi/h} {{ du_z} \over {2\pi}} \exp[-i(u_z z +
n\varphi)] \sum_{\omega,\omega^{\prime}}  \int_{-\pi/h}^{\pi/h} du_y \int_{-\pi/h}^{\pi/h} du_y^{\prime}
   P^*(\omega,{\bm u},n; {\bm r}) 
\nonumber \\ \times  
\langle \omega,u_y\vert M^{-1}(u_z,n)
   \vert \omega^{\prime},u_y^{\prime} \rangle
     \tilde{\psi}(\omega^{\prime},u_y^{\prime},u_z,n) \ .
\label{Eq6}
\end{eqnarray}

\noindent
Here

\begin{eqnarray}
&& \hspace{-1cm} P(\omega,{\bm u},n; {\bm r})=\sum_{k=-\infty}^{\infty}\sum_{\bm v}
a(\omega,{\bm u}+{\bm v},n+N_{\theta}k;R) 
\exp[i(N_{\theta}k\varphi+ v_z z)] \ ,
\label{P_a} \\
&& \hspace{-1cm} a(\omega,{\bm q},m;R) = 
\int_0^{2\pi}  \kappa(\omega,0,{\bm q};R\cos\varphi)  
\exp[i(q_yR\sin\varphi - m\varphi)] d\varphi \ ,
\label{a_kappa}
\end{eqnarray}

\noindent
the elements of the matrix $M(u_z,n)$ are given by 

\begin{eqnarray}
&& \hspace{-2cm}\langle \omega,u_y\vert M(u_z,n) \vert
\omega^{\prime},u_y^{\prime} \rangle = \sum_{k=-\infty}^{\infty}
\sum_{v_y,v_y^{\prime}} \sum_{v_z}
\langle \omega, u_y+v_y\vert
M_1(u_z+v_z,n+N_{\theta}k)\vert
\omega^{\prime},u_y^{\prime}+v_y^{\prime} \rangle \ , \nonumber \\
&& 
\label{M_M1} \\
&& \hspace{-2cm}\langle \omega,q_y\vert M_1(q_z,m)\vert \omega^{\prime},q_y^{\prime}
\rangle = \int_0^{L/2}
a(\omega,q_y,q_z,m;R) a^*(\omega^{\prime},q_y^{\prime},q_z,m;R) RdR
\label{M1}
\end{eqnarray}

\noindent
and the Fourier-transformed data function is

\begin{equation}
\label{Eq7}
\tilde{\psi}(\omega,{\bm u},n)=\sum_{{\bm \rho}_d,\theta}
\psi(\omega,{\bm \rho}_d,\theta) \exp[i({\bm u}\cdot{\bm \rho}_d +
n\theta)] \ .
\end{equation}

\noindent
Note that in (\ref{Eq7}) we have explicitly included the dependence of
the data function on the angle of orientation $\theta$. The functions
$a(\omega,{\bm q},m;R)$ and $\langle \omega,q_y\vert M_1(q_z,m)\vert
\omega^{\prime},q_y^{\prime} \rangle$ can be, in general, expressed in
terms of modified Bessel functions. The corresponding integrals
(\ref{a_kappa}) and (\ref{M1}) are calculated in the Appendix for the
case of purely absorbing boundaries.

A few comments on the reconstruction formula (\ref{Eq6}) are
necessary. First, there is an apparent difference between the
variables $u_z,n$ and $u_y,\omega$. The first set of variables
correspond (after Fourier transformation of the data) to the variables
$z,\theta$. These are the variables with respect to which the
unperturbed medium is translationally invariant, and they can be
referred to as ``external'' variables. The variables $\omega,u_y$ are
``internal'' variables: they do not correspond to any translational
invariance of the system.  Second, the reconstruction algorithm
(\ref{Eq7}) involves integration over the continuous variables $u_y$
and $u_y^{\prime}$ and inversion of the operator $M(u_z,n)$ whose
matrix elements depend on continuous indices.  However, if the
variables $u_y,u_y^{\prime}$ are discretized and the corresponding
integration in (\ref{Eq7}) is replaced by a summation, then $M(u_z,n)$
becomes a discrete matrix. The resulting reconstruction formula is no
longer an SVD pseudo-inverse on the whole set of data
$\psi(\omega,{\bm \rho}_d,\theta)$. However, it is a pseudo-inverse
solution on the set of the Fourier-transformed data
$\tilde{\psi}(\omega,u_y,u_z,\theta)$ where $u_y$ takes only discrete
values. Third, it can be verified that in the case $N_{\theta}=1$, the
reconstruction formula (\ref{Eq7}) reduces to (\ref{Eq4}). Fourth, we
note that the number of degrees of freedom in the data-function
$\tilde{\psi}$ is four ($\omega, u_y, u_z$ and $n$). Thus, when the
number of rotations is large, it is sufficient to use only one or a
few values of the variable $u_y$, in which case the inverse problem is
still well determined. It can be argued that the reconstruction
algorithm is then ``numerical'' in one dimension and ``analytic'' in
two.~\footnote{If the number of rotations and the number of discrete
  values of $u_y$ are both large, it should be possible to recover the
  absorption and scattering coefficients uniquely and simultaneously.
  This theoretical possibility is not discussed in this paper.}
However, when only a small number of projections is taken, we must use
a relatively large number of discrete values of $u_y$.  By doing so,
we increase the size of the matrix $M$ whose SVD must be found
numerically. The inverse solution (\ref{Eq6}) is then ``numerical'' in
two dimensions and ``analytic'' in one. A similar algorithm (numerical
in two dimensions and analytic in one dimension) was proposed and
implemented in~\cite{markel_04_2}, where the image reconstruction area
was rectangular rather than cylindrical, but only two orthogonal
projections were allowed. In contrast, the full potential of the image
reconstruction algorithm proposed here is realized when $N_{\theta}$
is large. 

\section{Numerical Results}
\label{sec:simulations}

\subsection{Single projection}
\label{singleproj_sim}

We have implemented the proposed reconstruction algorithm using
computer-generated data and the following parameters: the slab
thickness was chosen to be the same as the cw diffuse wavelength,
$L=2\pi\sqrt{D_0/\alpha_0}$ (for most biological tissues, this
corresponds to $L\sim 6{\rm cm}$); the lattice step was chosen to be
$h=L/40$ and we have used $N=25$ different modulation frequencies
which range from $\omega=0$ to $\omega=10\alpha_0$ (the maximum
frequency corresponds to $\sim 1.6{\rm GHz}$); the field of view was
chosen to be $L\times L$ and, finally, we have generated forward data
for a single point (delta-function) absorber which is located in the
center of the field of view but at different depths.  Absorbing
boundary conditions were imposed on the surface of the slab. The
corresponding expression for the function $\kappa(\omega, 0, {\bm
  q};x)$ is given in the Appendix.

\begin{figure}
\centerline{\psfig{file=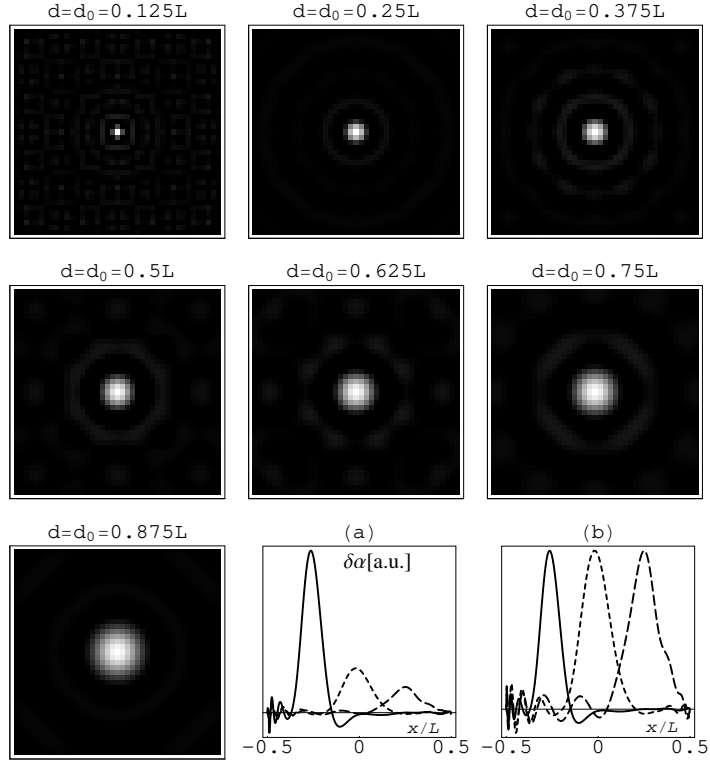,width=10cm}}
\caption{\label{fig:single_proj} 
  Tomographic slices parallel to the slab surface drawn through the
  medium at different depths $d$ (from the plane of scanned detection)
  with the small absorber lying in the center of the field of view at
  the same depth $d_0=d$, and the point-spread functions representing
  depth resolution (a,b). The curves are plotted on the same scale (a)
  and normalized to their own maxima (b). For curves (a,b), the point
  absorber depth is $d_0=0.25L$ (solid line), $d_0=0.5L$ (short dash)
  and $d_0=0.75L$ (long dash).}
\end{figure}

The results of reconstructions are shown in
Fig.~\ref{fig:single_proj}. The density plots represent tomographic
slices of the medium drawn at different depths $d$ (the distance from
the plane of scanned detection) parallel to the slab surfaces.  The
depth of the absorbing inhomogeneity, $d_0$, was in each case equal to
$d$; thus the slices represent the depth-dependent $y-z$ PSFs. Each
density plot has a linear color scale and is normalized to its own
maximum.  As expected, the PSFs become broader when the point absorber
approaches the illuminated plane. The last two panels (a,b) show the
PSFs in the depth direction ($x$) for point absorbers located at
$d_0=0.25L$, $d_0=0.5L$ and $d_0=0.75L$. Note that the approximate
half-widths of these curves are $0.06L$, $0.09L$ and $0.09L$,
respectively.

The analysis of Fig.~\ref{fig:single_proj} suggests that the PSFs are
depth-dependent. Moreover, the PSFs have different integral weights.
Thus, the point absorbers which are closer to the plane of scanned
detection result in higher peaks in the reconstructed images.  The
width of the PSFs also depends on depth of the point absorber.  This
potentially constitutes a serious problem for three dimensional
tomographic imaging.

\subsection{Multiple projections}
\label{subsec:multiproj_sim}

We have implemented numerically the multi-projection image
reconstruction formula (\ref{Eq6}).  Note that in the multi-projection
case there are two choices for graphically representing the
tomographic slices. In one case, the slices are perpendicular to the
axis of rotation. The image then is reconstructed in a circle. This
choice is convenient for studying the radial and angular resolutions.
Another possibility is to construct cylindrical slices $R=R_{\rm
  image}={\rm const}$, and map them onto rectangles. The image is then
reconstructed in the rectangular area $2\pi R_{\rm image} \times
(z_{\rm max}-z_{\rm min})$, where $z_{\rm max}$ and $z_{\rm min}$ are
the maximum and minimum values of $z$, chosen arbitrarily.

\begin{figure}
\centerline{\psfig{file=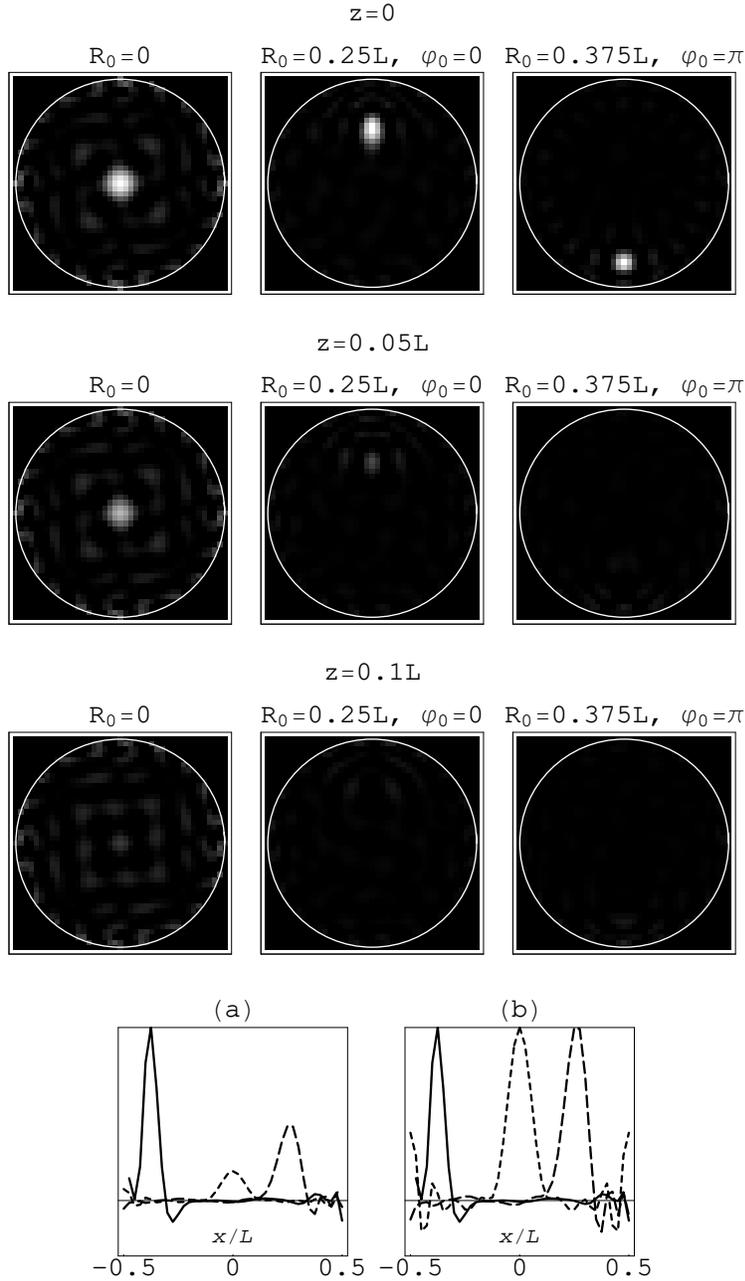,width=12.61cm}}
\caption{\label{fig:dbl_1}
  Circular slices illustrating radial, angular and $z$ resolution. All
  point absorbers are in the $z=0$ plane, and the point-spread
  functions representing depth ($R$) resolution (a,b). The radial and
  angular coordinates of the point absorber, $R_0$ and $\varphi_0$,
  are specified in the figure legends. First row of images: slices at
  $z=0$; second row: slices at $z=0.05L$; third row: slices at
  $z=0.1L$. Images (a-b): reconstruction along the diameter that
  crosses all three inhomogeneities. In (a,b) solid line corresponds
  to $R_0=0.375L$ and $\varphi_0=\pi$, short dash to $R_0=0$ and long
  dash to $R_0=0.25L$ and $\varphi_0=0$ Four projections, $15$
  modulation frequencies and $23$ discrete values of $u_y$ are used.}
\end{figure}

\begin{figure}
\centerline{\psfig{file=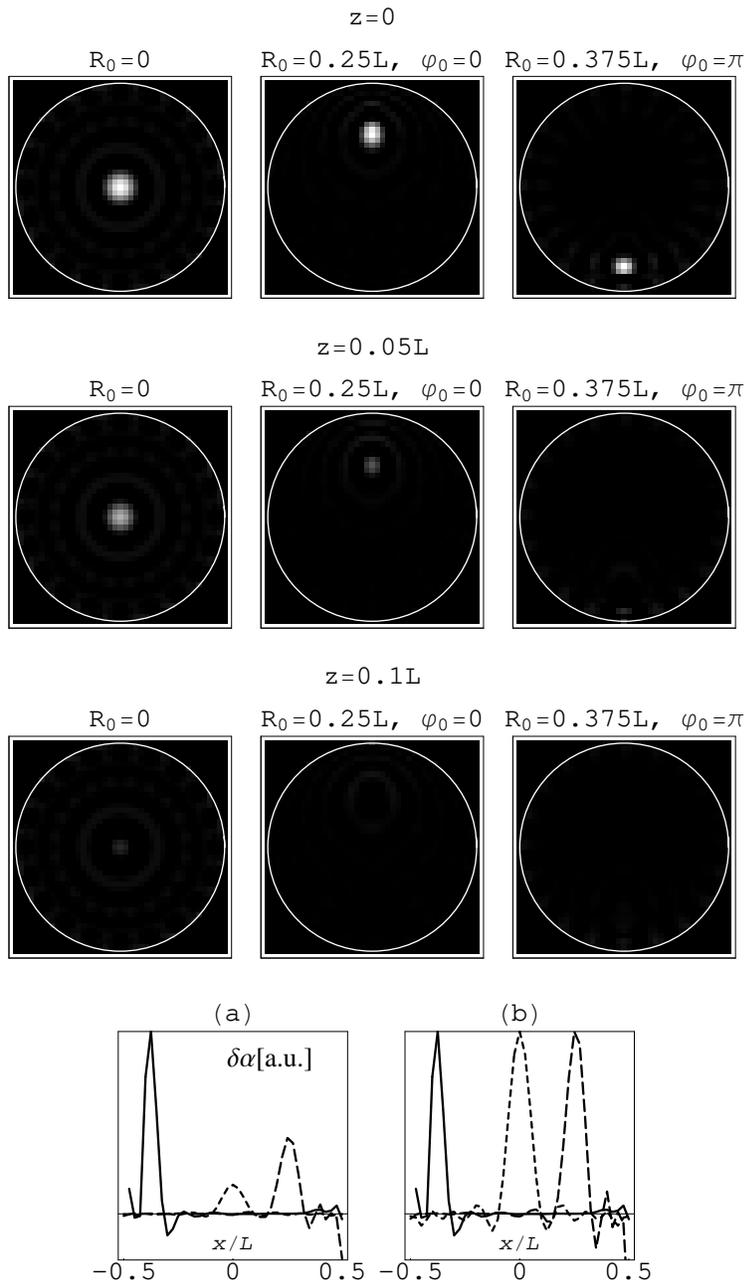,width=12.61cm}}
\caption{\label{fig:dbl_2} Same as in Fig.~\ref{fig:dbl_1} but $20$ projections are used.}
\end{figure}

\begin{figure}
\centerline{\psfig{file=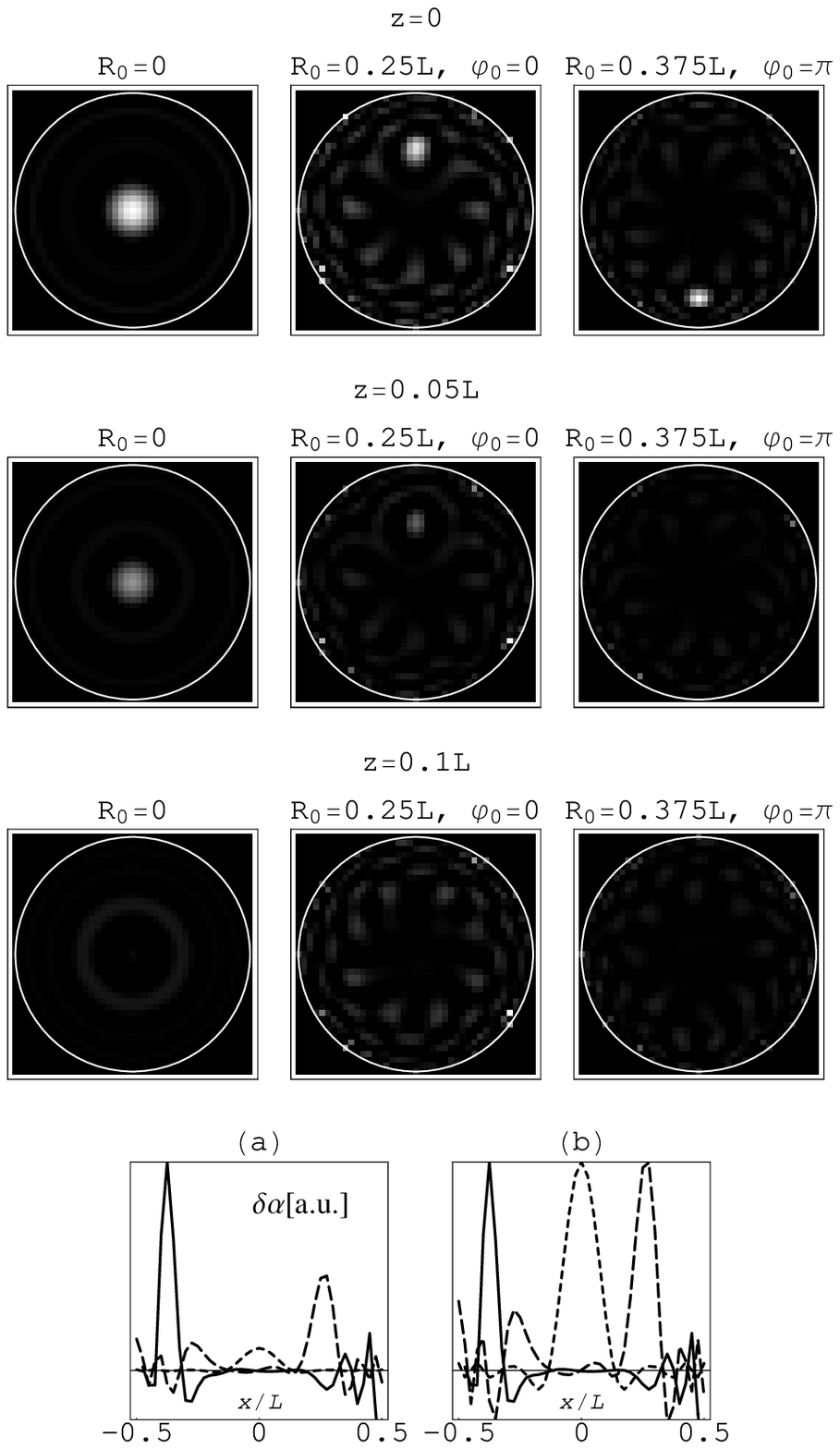,width=12.61cm}}
\caption{\label{fig:dbl_3} Same as in Fig.~\ref{fig:dbl_1} but $40$
  projections and only three discrete values of $u_y$ are used.}
\end{figure}

We start with the discussion of circular slices.  The results of
numerical implementation of the reconstruction formula (\ref{Eq7}) are
shown in Fig.~\ref{fig:dbl_1} for four different orientations of the
slab, namely $\theta=0,\pi/2,\pi,3\pi/2$.  We have used $23$ equally
spaced values of $u_y$ ranging from $-\pi/h$ to $\pi/h$ and $15$
equally spaced modulation frequencies ranging from $0$ to
$10\alpha_0$; otherwise, the parameters are the same as in
Fig.~\ref{fig:single_proj}. The inhomogeneity was located as specified
in the figure legend. The white spots in the density plots illustrate
the depth PSFs. The graphs (a,b) show the same PSFs in a more
quantitative way by plotting $\delta\alpha$ along the diameter of the
cylinder which intersects all three inhomogeneities.

As expected, using four different projections improves the image
quality by interchanging the source and detector planes, and the depth
and transverse directions. Moreover, using more projections than four
does not change the results substantially, as is illustrated in
Figs.~\ref{fig:dbl_2} and \ref{fig:dbl_3}. However, when a large
number of projections is taken, the inverse problem becomes well
determined even when a relatively small number of ``internal'' degrees
of freedom $u_y$ is used. This makes the reconstruction formulae
computationally efficient. Thus, the computation time required for
producing data for Fig.~\ref{fig:dbl_3} is more than an order of
magnitude less than that for Fig.~\ref{fig:dbl_1}, yet the image
quality is similar. We have verified that three discrete values of
$u_y$ is also sufficient for $N_\theta=20$ (taking a single value
$u_y=0$ results in a slight decrease in image quality; data not
shown).

Although the images shown in Figs.~\ref{fig:dbl_1}-\ref{fig:dbl_3} are
similar, the best image quality is, in fact, attained in
Fig.~\ref{fig:dbl_2}. Here the approximate half-widths of the PSF in
the $R$ direction are $0.05L$ for the inhomogeneity located at
$R_0=0$, $0.04L$ for the inhomogeneity at $R_0=0.25L$, $\varphi_0=0$;
and $0.03L$ for the inhomogeneity at $R_0=0.375L$, $\varphi_0=\pi$.
These values should be compared to the respective values given in the
discussion of Fig.~\ref{fig:single_proj}. In particular, the
inhomogeneity located at $h_0=0.5L$ in Fig.~\ref{fig:single_proj}
corresponds to the inhomogeneity at $R_0=0$ in
Figs.~\ref{fig:dbl_1}-\ref{fig:dbl_3} and is the most ``difficult'' to
reconstruct since it is located deep inside the medium.  It can be
seen that the PSF half-width in the image of this particular
inhomogeneity is reduced by approximately the factor of $2$ due to the
use of multiple projections.  In addition, the relative heights of the
maxima of the PSFs in Fig.~\ref{fig:dbl_1}-\ref{fig:dbl_3} do not
differ as much as in Fig.~\ref{fig:single_proj}. This is expected to
reduce image artifacts.

\begin{figure}
\centerline{\psfig{file=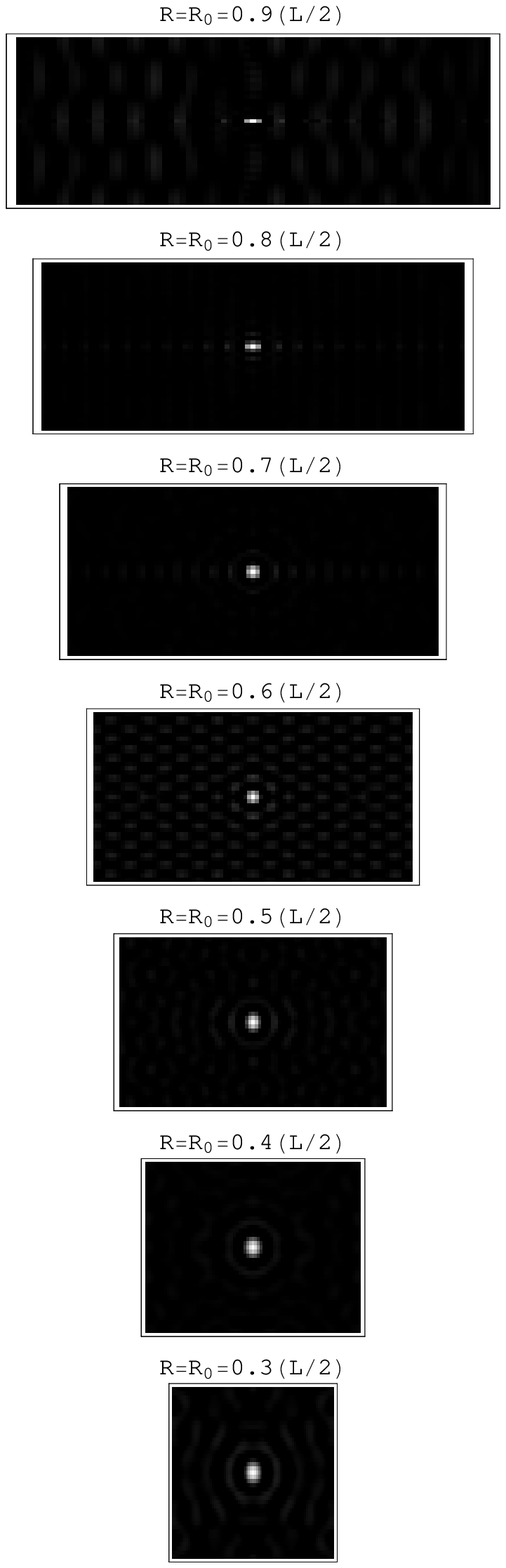,width=7.5cm}\psfig{file=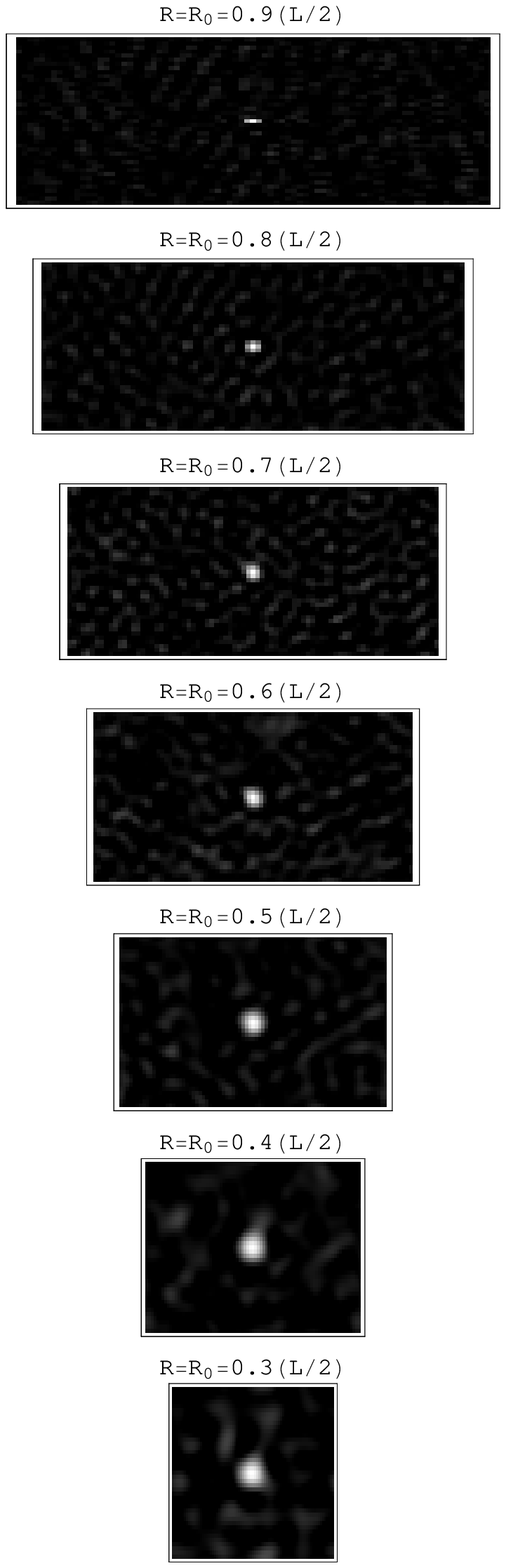,width=7.5cm}}
\centerline{(a) \hspace{7cm} (b)}
\caption{\label{fig:cyl_slices} 
  Cylindrical slices illustrating $z$ and $\varphi$-resolution for
  zero noise level (a) and for $1\%$ noise-to-signal ratio (b). The
  point absorbers are located in the $z=0$ plane at radial depths
  $R_0$ as indicated. The cylindrical surfaces with radii $R=R_0$
  (directly intersecting the inhomogeneity) are shown as projections
  onto a plane; the length of the vertical side of each rectangle is
  equal to $L$ and of the horizontal side to $2\pi R$. Forty
  projections, $25$ modulation frequencies and $9$ discrete values of
  $u_y$ are used for reconstruction.}
\end{figure}

Now we consider the cylindrical slices. From the computational point
of view, the use of cylindrical slices is a more natural way to
display reconstructed images. This is evident from the inversion
formulae (\ref{Eq6}),(\ref{P_a}). Indeed, it can be seen that when the
reconstructed image is rasterized so that the variables $z$ and
$\varphi$ are placed on lattices with steps $h$ and $2\pi/N_\theta$,
respectively, the function $P(\omega,{\bm u},n; R,z,\varphi)$ becomes
independent of $z$ and $\varphi$. Then the dependence of reconstructed
images on these two variables is only due to the exponent in the
integral (\ref{Eq6}) and the reconstruction formula, with respect to
these two variables, is reduced to a Fourier transform. In
Fig.~\ref{fig:cyl_slices} we have used three discrete values of $u_y$
with $40$ different projections and slices are drawn as described in
the figure caption. Fig.~\ref{fig:cyl_slices}(a) illustrates image
reconstruction with noiseless data. It can be directly compared to
slices shown in Fig.~\ref{fig:single_proj}.  To demonstrate the
stability of image reconstruction, we have added random Gaussian noise
to the data function at the level of 1\% of the average absolute value
of the data. The result is shown in Fig.~\ref{fig:cyl_slices}(b). As
is well known, inclusion of noise tends to decrease spatial
resolution. It can be seen that this effect is stronger for
inhomogeneities that are deeper inside the medium. We have
demonstrated earlier that multi-projection imaging is more stable in
the presence of noise than the single projection
technique~\cite{markel_04_2}.

\section{Summary}
\label{sec:summary}

In summary, we have presented a new experimental modality and
computationally efficient image reconstruction algorithms for optical
diffusion tomography employing plane wave illumination with multiple
projections. Note that due to reciprocity, plane wave illumination and
scanned detection is equivalent to illumination by a scanned narrow
beam and integrated detection (e.g., with the use of time-resolved CCD
camera). The following specific conclusions can be drawn

\begin{itemize}
  
\item Use of plane wave illumination may be simpler experimentally
  than the traditional approach in which point-like sources and
  detectors are scanned because measurements with a much smaller
  dynamic range are required.
  
\item In a single projection experiment, the image quality is
  relatively good when the point absorber is close to the scanned
  surface and deteriorates as it approaches the plane of illumination.
  This situation should be contrasted with the traditional point
  source/point detector modality~\cite{markel_02_2}, where the image
  quality is low for inhomogeneities located in the center of a slab
  and improves when the inhomogeneity approaches either of
  the imaging surfaces. For a point inhomogeneity in the center of a
  slab, the image quality is slightly better for the traditional (point
  source/point detector) modality (cf.~\cite{markel_02_2}).
  
\item Rotating the slab around the sample removes many of the
  deficiencies of the plane wave illumination scheme by
  interchanging the scanned and integrated detection surfaces and
  depth and transverse directions. A minimum of four projections is
  required for such an interchange.

\item When only four rotations are used, a large number of discrete
  values of the ``internal'' variable $u_y$ must be utilized in the
  reconstruction. Alternatively, a large number of projections can be
  used with a small number of discrete values of $u_y$. The second
  approach is much more computationally efficient but requires more
  complicated measurements. The quality of images is similar in both
  cases.
  
\item The plane wave illumination approach allows one to
  significantly simplify reconstruction formulae, both in single- and
  multiple-projection imaging.
  
\item If only small number of projections is used (two or four) an
  alternative approach may be used, which is purely numerical in two
  dimensions and analytic in one dimension~\cite{markel_04_2}. For a large
  number of projections, the algorithm reported here is
  computationally more efficient.

\end{itemize}

This work was supported in part by the AFOSR under the grant
F41624-02-1-7001 and by the NIH under grant P41RR0205.

\appendix
\section*{Appendix: Calculation of the functions $a(\omega,{\bm q},m;R)$
  and $M_1(q_z,m)$.}
\setcounter{equation}{0}
\renewcommand{\theequation}{A{\arabic{equation}}}

The function $a(\omega,{\bm q},m;R)$ is defined by (\ref{a_kappa}). To
evaluate the integral, we must specify the function $\kappa(\omega,
0,{\bm q}; x)$. Explicit expressions for $\kappa$ are given
in~\cite{markel_02_1} for general boundary conditions. In this paper
we consider absorbing boundaries for which $\kappa$ is given by the
expression

\begin{equation}
\label{a0}
\kappa(\omega,0,{\bm q};x) = \left( {\ell^* \over D_0} \right)^2 
{{\sinh[k(L/2-x)]\sinh[Q(L/2+x)]} \over {\sinh(kL)\sinh(QL)}}  \ ,
\end{equation}

\noindent
where $\ell^*=3D_0/c$ is the transport mean free path, $c$ is the
average speed of light in the medium,
$k=\sqrt{(\alpha_0-i\omega)/D_0}$ is the complex diffuse wavenumber,
$Q=\sqrt{q^2 + k^2}$ and ${\bm q}=(q_y,q_z)$.  Generalization to mixed
boundaries of Robin type is straightforward and is not discussed here.
Then, the expression for $a(\omega,{\bm q},m;R)$ becomes

\begin{eqnarray}
\label{a2}
&& \hspace{-2cm}a(\omega,{\bm q},m;R) = \left( {\ell^* \over D_0} \right)^2 { 1 \over
  {\sinh(kL)\sinh(QL)}} \nonumber \\
&& \hspace{-1.5cm} \times \int_0^{2\pi} \sinh[k(L/2
-R\cos\varphi)]\sinh[Q(L/2+R\cos\varphi)]
\exp[i(q_yR\sin\varphi-m\varphi)] d\varphi \ . \nonumber \\
\end{eqnarray}

\noindent
This can be equivalently rewritten as

\begin{eqnarray}
\label{a3}
&& \hspace{-2cm} a(\omega,{\bm q},m;R) = \left( {\ell^* \over D_0} \right)^2 { 1 \over
  {4 \sinh(kL)\sinh(QL)}} \nonumber \\
&& \hspace{-2cm} \times \biggl \{
     \exp\Bigl[( Q+k)L/2 \Bigr] F_m \Bigl [(Q-k)R, iq_y R \Bigr ]
-    \exp \Bigl [(-Q+k)L/2 \Bigr ] F_m \Bigl [ (-Q-k)R, iq_y R \Bigr ]
\biggr. \nonumber \\ 
&& \hspace{-2cm} - \biggl. \exp \Bigl [ ( Q-k)L/2 \Bigr ] F_m \Bigl [
( Q+k)R, iq_y R \Bigr ]
+    \exp \Bigl [ (-Q-k)L/2 \Bigr ] F_m \Bigl [ (-Q+k)R, iq_y R \Bigr ]
\biggr \} \nonumber \\
\end{eqnarray}

\noindent
where

\begin{eqnarray}
\label{a4}
\hspace{-2cm} F_m(u,v) = \int_0^{2\pi} \exp[u\cos\varphi + v\sin\varphi - im\varphi]
d \varphi =2\pi\left( { \sqrt{u^2 + v^2} \over {u + iv} }\right)^m I_m(\sqrt{u^2
  + v^2}) \ , \nonumber \\
\end{eqnarray}

\noindent
and $I_m(x)$ is the modified Bessel function of the first kind. Note
that (\ref{a4}) is well defined, including the case $v=iu$. 

The expressions (\ref{a3}) and (\ref{a4}) define $a({\bm q},m;R)$.
Next, we need to calculate the matrix elements of $M_1(q_z,m)$. This
integral contains sixteen terms of the form

\begin{eqnarray}
&& \hspace{-2cm} {{ s_1 s_2 s_3 s_4  \pi^2 (\ell^*/D_0)^2} \over
  {4\sinh(kL) \sinh(QL) \sinh(k^{\prime}L) \sinh(Q^{\prime}L)}}
\exp\left[(s_1 k + s_2 Q + s_3 k^{\prime} + s_4 Q^{\prime})L/2 \right]
\nonumber \\
&& \hspace{-2cm} \times \left[
 { { \sqrt{\left[ (-s_1 k + s_2 Q)^2 - q_y^2 \right] 
 \left[ (-s_3 k^{\prime} + s_4 Q^{\prime})^2 -
        (q_y^{\prime})^2\right]} } 
\over {(-s_1 k + s_2 Q - q_y) (-s_3
      k^{\prime} + s_4 Q^{\prime} - q_y^{\prime}) 
    } } \right]^m \nonumber \\
&& \hspace{-2cm} \times \int_0^{L/2} 
I_m\Bigl[ R \sqrt{(-s_1 k          + s_2 Q)^2          -  q_y^2 }
\Bigr ]
I_m\Bigl[ R \sqrt{(-s_3 k^{\prime} + s_4 Q^{\prime})^2 -
  (q_y^{\prime})^2} \Bigr] RdR \ .
\label{a5}
\end{eqnarray}

\noindent
where $s_k=\pm 1$, the sixteen terms correspond to sixteen possible
permutations of the signs of $s_k$ and the primed variables should be
understood as follows:
$k^{\prime}=\sqrt{(\alpha_0-i\omega^{\prime})/D_0}$ and $Q^{\prime} =
\sqrt{(q_y^{\prime})^2 + q_z^2 + (k^{\prime})^2 }$.  The integral in
(\ref{a5}) is evaluated with the use of

\begin{eqnarray}
\label{a6}
\hspace{-2cm} \int_0^c xI_n(ax) I_n(bx) dx = \left\{
\begin{array}{ll}
\frac{\displaystyle c}{\displaystyle a^2 - b^2} [aI_{n+1}(ac)I_n(bc) - bI_n(ac)I_{n+1}(bc)]
& \ , \ \ \ a \neq b \ ,\\
& \\
-{\displaystyle c^2 \over {\displaystyle 2}}
\left[I_n^{\prime}(ac)\right]^2 + {{\textstyle 1} \over {\textstyle 2}}\left(c^2 +
{{\displaystyle n^2} \over {\displaystyle a^2}}\right) I_n^2(ac) & \ , \ \ \ a = b \ .
\end{array}
\right.
\end{eqnarray}

\noindent
This completely defines all the functions necessary for implementation
of the multi-projection reconstruction algorithm.

\section*{References}

\bibliography{abbrevplain,article,tomography,book}

\end{document}